\documentclass[a4paper, 10pt]{article}
\usepackage{style/osameet2}
\renewcommand{\thesection}{\Roman{section}}
\renewcommand{\thesubsection}{\thesection.\Roman{subsection}}
\usepackage[utf8]{inputenc}
\usepackage{glossaries}
\usepackage[UKenglish]{babel}
\newcommand{\SetCapsType}{normalcaps}
\makeatletter
\providecommand{\SetCapsType}{smallcaps}

\long\def\@scTrue{smallcaps}
\long\def\@scFalse{normalcaps}
\newcommand{\acroSCaps}[1]{%
 \begingroup
  \ifx\SetCapsType\@scTrue 
    \textsc{#1}%
  \else
    \MakeUppercase{#1}%
  \fi
  \endgroup
}
\makeatother

\newcommand{\nAcronym}[4][]{%
	\newacronym[#1]{#2}{#3}{#4}
}

\makeatletter
\@ifpackageloaded{babel}{
    \newcommand{\usuk}[2]{%
        \iflanguage{USenglish}{#1}{#2}
    }
}{
    \newcommand{\usuk}[2]{%
        #1
    }
}
\makeatother

\usepackage[shortcuts]{extdash}% prevent line break after hyphen before QAM
\newcommand{\qam}[1]{%, % at end of line result in no spaces before and after command
    \ifglsused{QAM}%
        {#1\=/\gls{QAM}}%
        {#1\=/ary \gls{QAM}%
    }%
}%

\nAcronym{2A8PSK}{\acroSCaps{2a8psk}}{2-ary amplitude 8-ary phaseshift keying}

\nAcronym{4D}{\acroSCaps{4d}}{four-dimensional}
\nAcronym{4D64PRS}{\acroSCaps{4d-64prs}}{\usuk{four-dimensional 64-ary polarization-ring-switching}{four-dimensional 64-ary polarisation-ring-switching}}

\nAcronym{5B4D2A8PSK}{\acroSCaps{5b4d-2a8psk}}{5-bit four-dimensional two-amplitude 8-ary phase-shift keying}

\nAcronym{6B4D2A8PSK}{\acroSCaps{6b4d-2a8psk}}{6-bit four-dimensional two-amplitude 8-ary phase-shift keying}

\nAcronym{8D}{\acroSCaps{8d}}{eight-dimensional}\nAcronym{8D2048PRS}{\acroSCaps{8d-2048prs}}{eight-dimensional 2048-ary polarization-ring-switching}
\nAcronym{8D2048PRST1}{\acroSCaps{8d-2048prs-t1}}{eight-dimensional 2048-ary polarization-ring-switching type 1}
\nAcronym{8D2048PRST2}{\acroSCaps{8d-2048prs-t2}}{eight-dimensional 2048-ary polarization-ring-switching type 2}

\nAcronym{ABC}{\acroSCaps{abc}}{automatic bias controller}
\nAcronym{ADC}{\acroSCaps{adc}}{analog-to-digital converter}
\nAcronym{AIR}{\acroSCaps{air}}{achievable information rate}
\nAcronym{AOM}{\acroSCaps{aom}}{acoustic optical modulator}
\nAcronym{AR}{\acroSCaps{ar}}{achievable rate}
\nAcronym{ASE}{\acroSCaps{ase}}{amplified spontaneous emission}
\nAcronym{AWGN}{\acroSCaps{awgn}}{additive white Gaussian noise}

\nAcronym{BER}{\acroSCaps{ber}}{bit error rate}
\nAcronym{BICM}{\acroSCaps{bicm}}{bit-interleaved coded modulation}
\nAcronym{BMD}{\acroSCaps{bmd}}{bit-metric decoding}
\nAcronym{BPD}{\acroSCaps{bpd}}{balanced photo-diode}
\nAcronym{BPS}{\acroSCaps{bps}}{blind phase search}
\nAcronym{BRGC}{\acroSCaps{brgc}}{binary reflected Gray code}

\nAcronym{CCDM}{\acroSCaps{ccdm}}{constant composition distribution matching}
\nAcronym{CD}{\acroSCaps{cd}}{chromatic dispersion}
\nAcronym{CIR}{\acroSCaps{cir}}{channel impulse response}
\nAcronym{ChUT}{\acroSCaps{chut}}{channel under test}
\nAcronym{CUT}{\acroSCaps{cut}}{channel under test}
\nAcronym{CPE}{\acroSCaps{cpe}}{carrier phase estimation}
\nAcronym{CAGR}{\acroSCaps{cagr}}{compound annual growth rate}

\nAcronym{DA}{\acroSCaps{da}}{driver amplifier}
\nAcronym{DAC}{\acroSCaps{dac}}{digital-to-analog converter}
\nAcronym{DCF}{\acroSCaps{dcf}}{\usuk{dispersion compensated fiber}{dispersion compensated fibre}}
\nAcronym{DFB}{\acroSCaps{dfb}}{distributed feedback}
\nAcronym{DGD}{\acroSCaps{dgd}}{differential group delay}
\nAcronym{DM}{\acroSCaps{dm}}{distribution matcher}
\nAcronym{DMGD}{\acroSCaps{dmgd}}{differential mode group delay}
\nAcronym{DP}{\acroSCaps{dp}}{\usuk{dual-polarization}{dual-polarisation}}
\nAcronym{DPC}{\acroSCaps{dpc}}{digital pre-compensation}
\nAcronym{DPE}{\acroSCaps{dpe}}{digital pre-emphasis}
\nAcronym{DPIQ}{\acroSCaps{dp-iqm}}{\usuk{dual-polarization IQ-modulator}{dual-polarisation IQ-modulator}}
\nAcronym{DRE}{\acroSCaps{dre}}{digital resolution enhancer}
\nAcronym{DSO}{\acroSCaps{dso}}{digital sampling oscilloscope}
\nAcronym{DSP}{\acroSCaps{dsp}}{digital signal processing}
\nAcronym{DUT}{\acroSCaps{dut}}{device under test}
\nAcronym{DWDM}{\acroSCaps{dwdm}}{dense wavelength-division multiplexing}

\nAcronym{ECL}{\acroSCaps{ecl}}{external cavity laser}
\nAcronym{ED}{\acroSCaps{ed}}{Eucledian distance}
\nAcronym{EDFA}{\acroSCaps{edfa}}{\usuk{erbium doped fiber amplifier}{erbium doped fibre amplifier}}
\nAcronym{ENOB}{\acroSCaps{enob}}{effective number of bits}
\nAcronym{ESS}{\acroSCaps{ess}}{enumerative sphere shaping}

\nAcronym{FDE}{\acroSCaps{fde}}{\usuk{frequency domain equalizer}{frequency domain equaliser}}
\nAcronym{FEC}{\acroSCaps{fec}}{forward error correction}
\nAcronym{FFT}{\acroSCaps{fft}}{fast Fourier transform}
\nAcronym{FMF}{\acroSCaps{fmf}}{\usuk{few-mode fiber}{few-mode fibre}}
\nAcronym[plural=FM-MCF, firstplural=\usuk{few-mode multi-core fibers}{few-mode multi-core fibres}]{FM-MCF}{\acroSCaps{fm-mcf}}{\usuk{few-mode multi-core fiber}{few-mode multi-core fibre}}
\nAcronym{FWM}{\acroSCaps{fwm}}{four-wave mixing}

\nAcronym{GFF}{\acroSCaps{gff}}{gain flattening filter}
\nAcronym{GMI}{\acroSCaps{gmi}}{\usuk{generalized mutual information}{generalised mutual information}}
\nAcronym{GS}{\acroSCaps{gs}}{geometric shaping}
\nAcronym{GV}{\acroSCaps{gv}}{group-velocity}
\nAcronym{GVD}{\acroSCaps{gvd}}{group-velocity dispersion}

\nAcronym{HDFEC}{\acroSCaps{hd-fec}}{hard decision forward error correction}

\nAcronym[plural=IL, firstplural=insertion losses (\textsc{il})]{IL}{\acroSCaps{il}}{insertion loss}
\nAcronym{IFFT}{\acroSCaps{ifft}}{inverse fast Fourier transform}
\nAcronym{ISI}{\acroSCaps{isi}}{intersymbol interference}

\nAcronym{KK}{\acroSCaps{kk}}{Kramers-Kronig}

\nAcronym{LCOS}{\acroSCaps{LCoS}}{liquid crystal on silicon}
\nAcronym{LDPC}{\acroSCaps{ldpc}}{low-density parity-check}
\nAcronym{LEAF}{\acroSCaps{leaf}}{\usuk{large effective area fiber}{large effective area fibre}}
\nAcronym{LMS}{\acroSCaps{lmf}}{least means square}
\nAcronym{LLR}{\acroSCaps{llr}}{log-likelihood ratio}
\nAcronym{LO}{\acroSCaps{lo}}{local oscillator}
\nAcronym{LP}{\acroSCaps{lp}}{\usuk{linear polarized}{linear polarised}}
\nAcronym{LSPS}{\acroSCaps{lsps}}{\usuk{loop-synchronized polarization scrambler}{loop-synchronised polarisation scrambler}}

\nAcronym{MB}{\acroSCaps{mb}}{Maxwell-Bolzmann}
\nAcronym{MCF}{\acroSCaps{mcf}}{\usuk{multi-core fiber}{multi-core fibre}}
\nAcronym{MDG}{\acroSCaps{mdg}}{mode dependent gain}
\nAcronym[plural=MDL, firstplural=mode dependent losses (\textsc{mdl})]{MDL}{\acroSCaps{mdl}}{mode dependent loss}
\nAcronym{MDM}{\acroSCaps{mdm}}{mode division multiplexing}
\nAcronym{MF}{\acroSCaps{mf}}{matched filter}
\nAcronym{MI}{\acroSCaps{mi}}{mutual information}
\nAcronym{MIMO}{\acroSCaps{mimo}}{multiple-input multiple-output}
\nAcronym{MMA}{\acroSCaps{mma}}{multi-modulus algorithm}
\nAcronym{MMF}{\acroSCaps{mmf}}{\usuk{multi-mode fiber}{multi-mode fibre}}
\nAcronym{MMSE}{\acroSCaps{mmse}}{minimum mean squared error}
\nAcronym{MPLC}{\acroSCaps{mplc}}{multi-plane light converter}
\nAcronym{MSE}{\acroSCaps{mse}}{mean squared error}
\nAcronym{MUX}{\acroSCaps{mux}}{multiplexer}
\nAcronym{MZM}{\acroSCaps{mzm}}{Mach-Zehnder modulator}

\nAcronym{NF}{\acroSCaps{nf}}{noise figure}
\nAcronym{NGMI}{\acroSCaps{ngmi}}{\usuk{normalized generalized mutual information}{normalised generalised mutual information}}
\nAcronym{NIR}{\acroSCaps{nir}}{near infra-red}

\nAcronym{OFDR}{\acroSCaps{ofdr}}{optical frequency-domain reflectometer}
\nAcronym{OMFT}{\acroSCaps{omft}}{optical-multi-format transmitter}
\nAcronym{OSA}{\acroSCaps{osa}}{\usuk{optical spectrum analyzer}{optical spectrum analyser}}
\nAcronym{OSNR}{\acroSCaps{osnr}}{optical signal-to-noise ratio}
\nAcronym{OTDR}{\acroSCaps{otdr}}{optical time-domain reflectometer}
\nAcronym{OTF}{\acroSCaps{otf}}{optical tunable filter}
\nAcronym{OVNA}{\acroSCaps{ovna}}{\usuk{optical vector network analyzer}{optical vector network analyser}}

\nAcronym{PAM}{\acroSCaps{pam}}{pulse-amplitude modulation}
\nAcronym{PAS}{\acroSCaps{pas}}{probabilistic amplitude shaping}
\nAcronym{PAPR}{\acroSCaps{papr}}{peak-to-avarage power ratio}
\nAcronym{PBS}{\acroSCaps{pbs}}{polarization beam splitter}
\nAcronym{PCVD}{\acroSCaps{pcvd}}{plasma chemical vapor depostion}
\nAcronym{PDL}{\acroSCaps{pdl}}{polarization dependent loss}
\nAcronym{PL}{\acroSCaps{pl}}{photonic lantern}
\nAcronym{PMBPSK}{\acroSCaps{pm-bpsk}}{polarization-multiplexed binary phase-shift-keying}
\nAcronym{PMQPSK}{\acroSCaps{pm-qpsk}}{polarization-multiplexed quaternary phase-shift-keying}
\nAcronym{PM8QAM}{\acroSCaps{pm-8qam}}{polarization-multiplexed 8-ary quadrature amplitude modulation}
\nAcronym{PMD}{\acroSCaps{pmd}}{polarization mode dispersion}
\nAcronym{PNOB}{\acroSCaps{pnob}}{physical number of bits}
\nAcronym{PRBS}{\acroSCaps{prbs}}{pseudorandom bit sequence}
\nAcronym{PS}{\acroSCaps{ps}}{probabilistic shaping}

\nAcronym{QAM}{\acroSCaps{qam}}{quadrature amplitude modulation}
\nAcronym{QPSK}{\acroSCaps{qpsk}}{quadrature phase shift keying}

\nAcronym{RRC}{\acroSCaps{rrc}}{root-raised-cosine}

\nAcronym{SamPerSym}{\acroSCaps{sps}}{samples per symbol}
\nAcronym{SDFEC}{\acroSCaps{sd-fec}}{soft decision forward error correction}
\nAcronym{SDM}{\acroSCaps{sdm}}{space-division multiplexing}
\nAcronym{SE}{\acroSCaps{se}}{spectral efficiency}
\nAcronym{SER}{\acroSCaps{ser}}{symbol error rate}
\nAcronym{SA}{\acroSCaps{sa}}{simulated annealing}
\nAcronym{SLM}{\acroSCaps{slm}}{spatial light modulator}
\nAcronym{SSFM}{\acroSCaps{ssfm}}{split-step Fourier method}
\nAcronym{SMF}{\acroSCaps{smf}}{\usuk{single-mode fiber}{single-mode fibre}}
\nAcronym{SNR}{\acroSCaps{snr}}{signal-to-noise ratio}
\nAcronym[firstplural=\usuk{states of polarization (\acroSCaps{sop})}{states of polarisation (\acroSCaps{sop})}]{SOP}{\acroSCaps{sop}}{\usuk{state of polarization}{state of polarisation}}
\nAcronym{SPM}{\acroSCaps{spm}}{self-phase modulation}
\nAcronym{SPS}{\acroSCaps{sps}}{\usuk{synchronized polarization scrambler}{synchronised polarisation scrambler}}
\nAcronym{SSMF}{\acroSCaps{ssmf}}{\usuk{standard single-mode fiber}{standard single-mode fibre}}
\nAcronym{SVD}{\acroSCaps{svd}}{singular value decomposition}
\nAcronym{SWI}{\acroSCaps{swi}}{swept wavelength interferometry}

\nAcronym{TDE}{\acroSCaps{tde}}{\usuk{time domain equalizer}{time domain equaliser}}
\nAcronym{TDMSDM}{\acroSCaps{tdm-sdm}}{time-domain multiplexed space-division multiplexing}
\nAcronym{TE}{\acroSCaps{TE}}{transverse electric}
\nAcronym{TM}{\acroSCaps{TM}}{transverse magnetic}
\nAcronym{TH4D}{\acroSCaps{th-4d}}{time domain hybrid four-dimensional}
\nAcronym{TH4D2A8PSK}{\acroSCaps{th-4d-2a8psk}}{time domain hybrid four-dimensional two-amplitude eight-phase shift keying}

\nAcronym{VOA}{\acroSCaps{voa}}{variable optical attenuator}

\nAcronym{WDM}{\acroSCaps{wdm}}{wavelength division multiplexing}
\nAcronym{WSS}{\acroSCaps{wss}}{wavelength selective switch}

\nAcronym{XPM}{\acroSCaps{xpm}}{cross-phase modulation}

\nAcronym{3DWG}{\acroSCaps{3dwg}}{3D-waveguide}
\usepackage[capitalise]{cleveref}
\usepackage{adjustbox}
\usepackage{bm}
\usepackage[per-mode=symbol]{siunitx}
\usepackage[font=footnotesize]{subcaption}
\usepackage{booktabs}
\usepackage{floatrow}
\usepackage{amsmath,amssymb}
\usepackage{newtxmath}

\begin{document}
\title{First Experimental Demonstration of Probabilistic Enumerative Sphere Shaping in Optical Fiber Communications}
\vspace{-4mm}

\author{
Sebastiaan Goossens\textsuperscript{1},
Sjoerd van der Heide\textsuperscript{1},
Menno van den Hout\textsuperscript{1},
Abdelkerim Amari\textsuperscript{1},
Yunus Can Gültekin\textsuperscript{1},
Olga Vassilieva\textsuperscript{2},
Inwoong Kim\textsuperscript{2},
Tadashi Ikeuchi\textsuperscript{2},
Frans M. J. Willems\textsuperscript{1},
Alex Alvarado\textsuperscript{1},
and Chigo Okonkwo\textsuperscript{1}
}

\address{\textsuperscript{1}{Department of Electrical Engineering, Eindhoven University of Technology, The Netherlands} \\
\textsuperscript{2}{Fujitsu Laboratories of America, Inc., 2801 Telecom Parkway, Richardson TX 75082, USA} \\
s.a.r.goossens@student.tue.nl}

\vspace{-5mm}

\begin{abstract}
We transmit probabilistic enumerative sphere shaped dual-polarization 64-QAM at 350Gbit/s/channel over 1610km SSMF using a short blocklength of 200. A reach increase of 15\% over constant composition distribution matching with identical blocklength is demonstrated.
\end{abstract}
\keywords{Advanced Modulation, Coding and Multiplexing. Coding and forward error correction for optical communications.}

\maketitle
\vspace{-2mm}
\section{Introduction}
Fiber optical transmission systems are approaching their capacity limits~\cite{Winzer2017}. In recent years, several techniques have been proposed for increasing spectral efficiency in order to keep up with traffic demands. Within this context, constellation shaping in general---and \gls{PS} in particular---have received considerable interest. \Gls{PS} uses  uniformly spaced constellation points occurring with different probabilities, which can theoretically provide shaping gains up to 1.53 dB \glsentrylong{SNR} for the \glsentrylong{AWGN} channel~\cite{Forney1984}. Even higher gains for the nonlinear fiber optical channel have been reported \cite{Dar2014}. Consequently, the development of constellation shaping algorithms integrated with coded modulation has been the subject of intense research~\cite{Bocherer2015,Fehenberger2019,Buchali2016,Fehenberger2016}.

\begin{figure}\RawFloats\TopFloatBoxes
\vspace*{-6mm}
\begin{floatrow}
\hspace{-4mm}
\ffigbox{%
      \centering
    \begin{adjustbox}{width=0.95\linewidth}
	\includegraphics{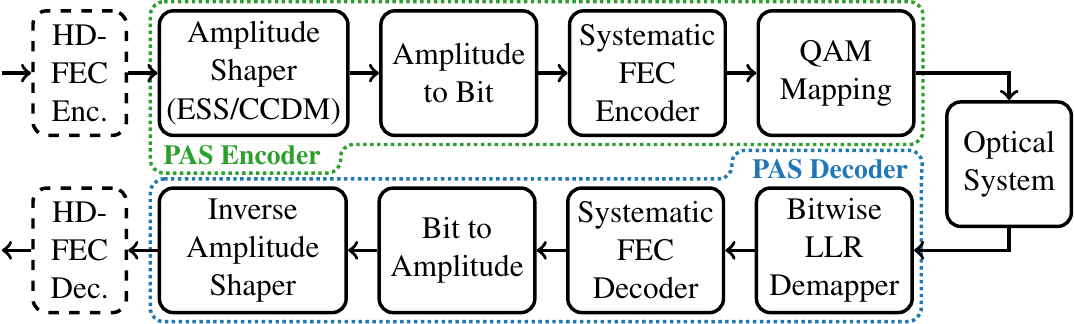}
    \end{adjustbox}
}{%
    \caption{Block diagram of the PAS system under consideration.}
    \label{fig:shaping_block_diagram}
}
\ttabbox{%
  \caption{Parameters for the four systems under consideration.}%
  \label{tab:shaping_parameters}
}{%
\def\arraystretch{1.0}%
\footnotesize
\hspace{-0.3cm}
\begin{tabular}{p{1.49cm} cccc}
	\toprule
	\bf{Name} & \bf{Shaping} & \bf{Block length} & \bf{FEC rate} & \bf{Net data rate} \\
	\midrule
	Uniform   & -       & -            & 3/4       & 9 bit/4D-sym \\
	CCDM-200  & CCDM    & 200          & 4/5       & 9 bit/4D-sym \\
	CCDM-3600 & CCDM    & 3600         & 4/5       & 9 bit/4D-sym \\
	ESS-200   & ESS     & 200          & 4/5       & 9 bit/4D-sym \\
	\bottomrule
\end{tabular}
}
\end{floatrow}
\vspace{-5mm}
\end{figure}

\Gls{PAS} \cite{Bocherer2015} integrates a shaping algorithm into an existing \gls{BICM} system \cite{AlexBICMBook}, as shown in \cref{fig:shaping_block_diagram}. \Gls{CCDM} was introduced in \cite{Bocherer2015} and is one of the most popular ways of implementing \gls{PAS}. \gls{CCDM} is based on a \gls{DM} which requires long blocklengths in order to reach optimum performance~\cite{Schulte2016}. Short blocklengths can be used to reduce the implementation challenges associated with the required arithmetic coding, however, this results in relatively large rate losses~\cite{Fehenberger2019}.

In this paper, an alternative probabilistic shaping algorithm based on \gls{ESS} is experimentally validated for the first time. \gls{ESS} was introduced in 1993 \cite{Willems1993} and has been recently considered for wireless communications \cite{Gultekin2019}. Very recently, \gls{ESS} has been introduced to the optical community in \cite{Amari2019}. The main result of \cite{Amari2019} is that short blocklength ESS outperforms CCDM at the same blocklength, long blocklength CCDM, and uniform. This gain is due to the combination of linear shaping gain and nonlinear tolerance. The results in \cite{Amari2019}, however, are only based on numerical (split-step Fourier) simulations.

In this paper, \qam{64}-based \gls{ESS} is experimentally compared against both \gls{CCDM} and uniform signalling, following the same setup considered in \cite{Amari2019}. Both \gls{CCDM} and \gls{ESS} indicate good performance with respect to the baseline uniform \gls{QAM}. However, \gls{ESS} is demonstrated to outperform \gls{CCDM} after long-haul optical transmission at equal blocklengths. For a net rate of 350 Gbit/s per channel, gains of 15\% translating into \SI{210}{\kilo\meter} reach increase are demonstrated for \gls{ESS} with respect to \gls{CCDM} at a blocklength of 200.

\vspace{-2mm}
\section{PAS with CCDM and ESS}

An amplitude shaper transforms a sequence of \(k\) uniformly distributed input bits into a block of nonuniformly distributed output amplitudes with blocklength \(n\). The shaping rate \(R_\text{s}\) is thus \(R_\text{s}={k}/{n}\). Finite blocklength shaping algorithms suffer from a rate loss which decreases with increasing blocklength~\cite{Schulte2016}. This rate loss is defined as \(R_\text{loss} = H(A) - R_\text{s}\), where \(H(A)\) is the entropy of the shaped amplitudes. As the complexity of shaping algorithms is generally linked to blocklength, it is clear that \gls{PS} algorithms with both short blocklength and low rate loss are not trivial to devise.

Based on arithmetic coding, \gls{CCDM} maps an input bitstream to any desired output amplitude distribution provided the input bits have equal probabilities. The main property of \gls{CCDM} is that every shaped block contains the same empirical set of amplitudes, and thus, satisfies the desired amplitude distribution per block. Therefore, it is only able to address a relatively limited set of sequences, and thus, is heavily impacted by rate loss at short blocklengths. \gls{ESS}, on the other hand, is a shaping algorithm based on sphere coding which \emph{indirectly} induces a nonuniform distribution. In \gls{ESS}, the output amplitude sequences are bounded-energy sequences, i.e., all sequences satisfy a maximum energy constraint \cite{Gultekin2019,Amari2019}. As a consequence of this, more input bits \(k\) are used for the same blocklength \(n\), resulting in a much lower rate loss. Using a geometric analogy to further explain the differences between the schemes, CCDM addresses sequences on the shell of a $n$-sphere, while ESS addresses all sequences in the CCDM shell as well as all inner shells (with lower energy).

\begin{figure*}[!b]
\vspace{-3mm}
    \begin{minipage}[c]{0.815\linewidth}
        \centering
        \includegraphics[width=\linewidth]{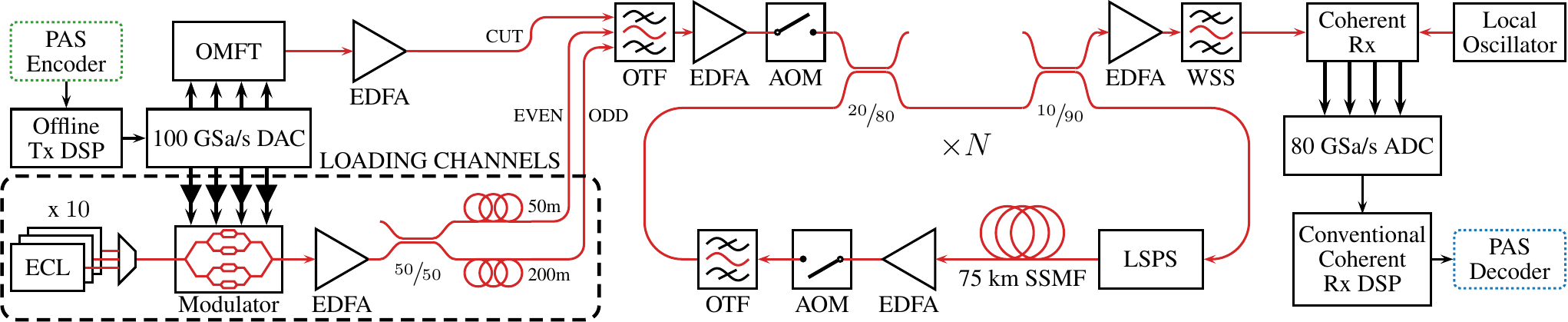}
    \end{minipage}
    \begin{minipage}[c]{0.18\linewidth}
    \begin{subfigure}[b]{0.485\linewidth}
        \centering
        \includegraphics[width=0.8\linewidth]{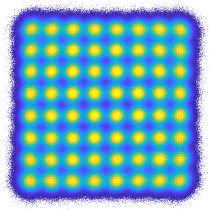}
        \vspace{-1.5ex}
        \captionsetup{font=scriptsize}
        \caption*{Uniform}
    \end{subfigure}
    \begin{subfigure}[b]{0.485\linewidth}
        \centering
        \includegraphics[width=0.8\linewidth]{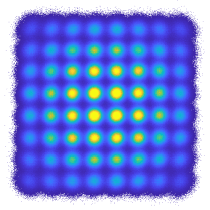}
        \vspace{-1.5ex}
        \captionsetup{font=scriptsize}
        \caption*{CCDM-200}
    \end{subfigure}
    
    \begin{subfigure}[b]{0.485\linewidth}
        \centering
        \includegraphics[width=0.8\linewidth]{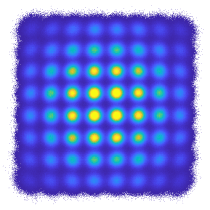}
        \vspace{-1.5ex}
        \captionsetup{font=scriptsize}
        \caption*{CCDM-3600}
    \end{subfigure}
    \begin{subfigure}[b]{0.485\linewidth}
        \centering
        \includegraphics[width=0.8\linewidth]{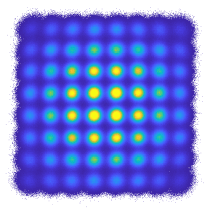}
        \vspace{-1.5ex}
        \captionsetup{font=scriptsize}
        \caption*{ESS-200}
    \end{subfigure}
    \end{minipage}
    
    \caption{Experimental optical recirculating loop setup (left) and received constellations (right).}
    \label{fig:ECOC19_setup}
\vspace{-5mm}
\end{figure*}

\vspace{-2mm}
\section{Experimental Setup}

In this paper, all signals are based on \qam{64}. As a baseline, uniform signaling is used with a \gls{FEC} rate of 3/4, resulting in a net information rate of 9 bit/4D-sym. For PAS, an \gls{FEC} rate of 4/5 is employed. The constellation is subsequently shaped to an entropy such that the net information rate is equal to 9 bit/4D-sym (accounting for rate loss). Three shaped schemes are considered: ESS and CCDM with a blocklength of 200, and CCDM with a blocklength of \SI{3600} (a case with negligible rate loss). A summary is shown in Table~\ref{tab:shaping_parameters}. For the experiment, 20 \gls{LDPC} blocks of shaped amplitudes are constructed, which are then encoded using the DVB-S2 \gls{LDPC} code. These bits are mapped to their corresponding symbols following the \gls{PAS} encoder in \cref{fig:shaping_block_diagram}.

The sequences are generated offline and contain \SI{216000} 64-\gls{QAM} symbols, which are pulse-shaped using a \gls{RRC} filter with 1\% roll-off at 41.79 GBd and uploaded to a 100-GSa/s \gls{DAC}, as shown in \cref{fig:ECOC19_setup} (left). The 1550.116~nm \gls{CUT} is modulated using an \gls{OMFT}, which consists of an \gls{ECL}, a \gls{DPIQ}, an \gls{ABC} and RF-amplifiers. The multiplexed outputs of 10 \glspl{ECL} are modulated using a \gls{DPIQ}, amplified, split into odd and even channels before being decorrelated by \SI{10200} symbols (50~m) and \SI{40800} (200~m) with respect to the \gls{CUT}, respectively. The \gls{CUT}, odd and even channels are combined onto the 50-GHz spaced \gls{DWDM} grid using an \gls{OTF}. Using \glspl{AOM}, the signal is circulated in a loop consisting of a \gls{LSPS}, a 75-km span of \gls{SSMF}, an \gls{EDFA}, and an \gls{OTF} used for gain equalisation. After transmission, the signal is amplified, filtered using a \gls{WSS}, detected using an intradyne coherent receiver and digitized by an 80-GSa/s real-time oscilloscope. The receiver \gls{DSP} is performed offline and consists of front-end compensation, \glsentrylong{CD} compensation, frequency-offset compensation, and equalization with in-loop phase correction. To reduce the influence from sub-optimal receiver \gls{DSP} implementation and stability, such as \gls{SER} influence on equalizer feedback performance, training symbols are used for all processed data. On average, 50 sequences have been captured with the oscilloscope resulting in a total of \SI{1000} received \gls{LDPC} blocks per launch power setting and per distance.

After the conventional \gls{DSP}, the received symbols are demodulated using a soft-decision bit-wise demapper taking into account the probabilities per symbol. The \glsentrylong{LLR} values are then fed into the \gls{LDPC} decoder with a maximum of 50 decoding iterations. If shaping is considered, these bits are further processed via the \gls{PAS} decoder (see \cref{fig:shaping_block_diagram}). \cref{fig:ECOC19_setup}~(right) shows the received constellations after transmission over \SI{375}{\kilo\meter} for the four cases outlined in Table~\ref{tab:shaping_parameters}.

\vspace{-2mm}
\section{Results}
The results are shown in terms of end-to-end \gls{BER} and \gls{AIR} for finite length \glspl{BMD} with PAS. The latter is defined as (in \cite[eq.~(26)]{Fehenberger2019})
 \begin{equation}\label{gmi}
     \text{AIR}_{n} = \underbrace{\left[  H(\boldsymbol{C}) - \sum_{i=1}^6 H(C_i \mid Y) \right]}_{\text{BMD Rate}} - \underbrace{\left[H(A)-\frac{k}{n}\right]}_{\text{Rate loss}},
 \end{equation}
 where $\boldsymbol{C}=(C_1,C_2,\ldots,C_6)$ are the bits at the input of the mapper, and $Y$ the received symbols after all DSP.

\begin{figure}
	\vspace{-2mm}
	\centering
	\begin{subfigure}[t]{.45\linewidth}%
		\centering
		\includegraphics{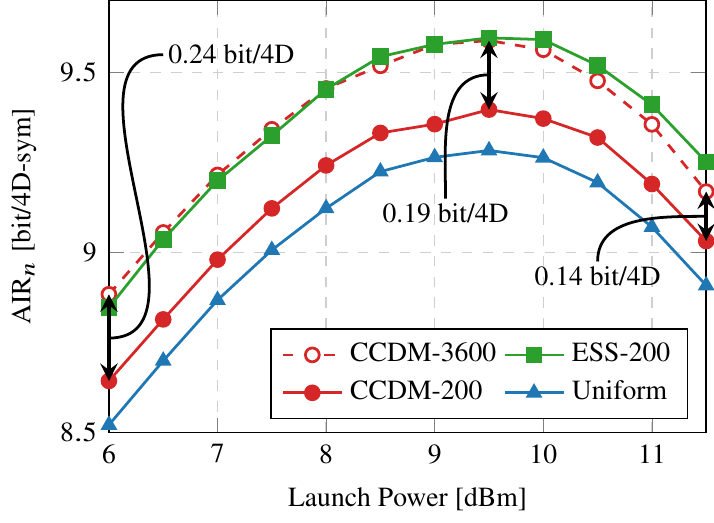}
		\vspace{-2mm}%
		\caption{}%
		\label{fig:LPswp_GMI_Rloss}%
		\vspace{-3mm}%
	\end{subfigure}%
	\begin{subfigure}[t]{.45\linewidth}%
		\centering
		\includegraphics{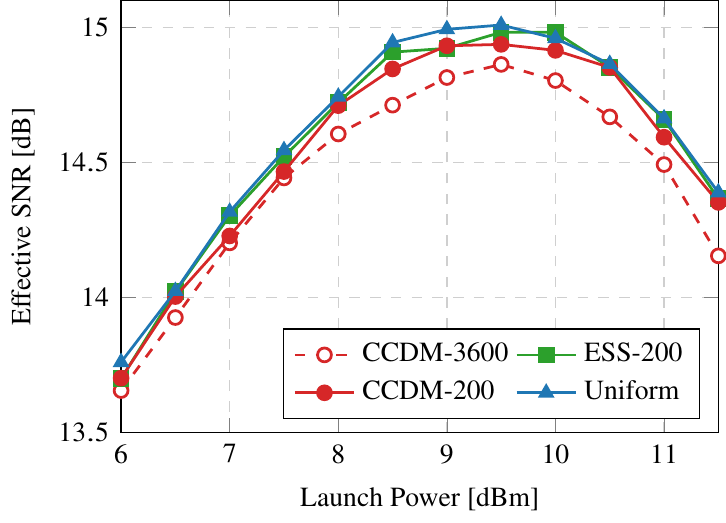}
		\vspace{-2mm}%
		\caption{}%
		\label{fig:LPswp_SNReff}%
		\vspace{-3mm}%
	\end{subfigure}
	\begin{subfigure}[t]{.45\linewidth}%
		\centering
		\includegraphics{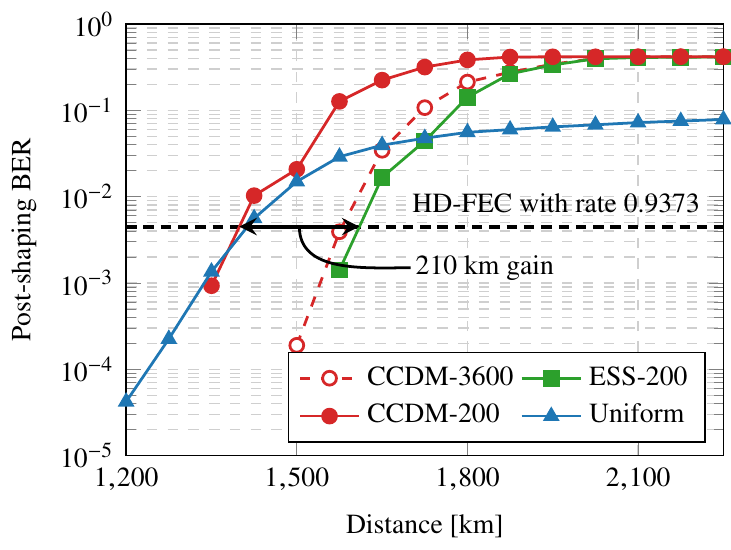}
		\vspace{-2mm}%
		\caption{}%
		\label{fig:Dswp_BER_post_DM}%
		\vspace{-4mm}%
	\end{subfigure}%
	\begin{subfigure}[t]{.45\linewidth}%
		\centering
		\includegraphics{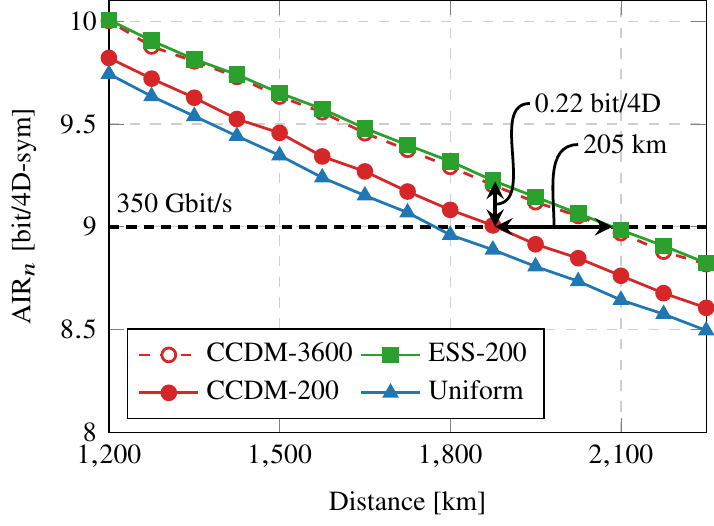}
		\vspace{-2mm}%
		\caption{}%
		\label{fig:Dswp_GMI_Rloss}%
		\vspace{-4mm}%
	\end{subfigure}
	\caption{(a) AIR\(_n\) versus launch power at \(1500\) km (\(20\) spans). \,\,\,(b) Effective SNR versus launch power at \(1500\) km (\(20\) spans). \,\,\,(c) BER after inverse amplitude shaper versus transmission distance at the optimal launch power of \(9.5\) dBm. \,\,\,(d) AIR\(_n\) versus transmission distance at the optimal launch power of \(9.5\) dBm.\vspace{-5mm}}
	
\end{figure}

\cref{fig:LPswp_GMI_Rloss} shows the \gls{AIR} vs. launch power for a transmission distance of \SI{1500}{\kilo\meter} and the four cases under consideration. In the linear regime (low input power), \cref{fig:LPswp_GMI_Rloss} shows that long blocklength \gls{CCDM}-3600 outperforms \gls{ESS}-200 and \gls{CCDM}-200. In the nonlinear regime, \gls{ESS}-200 exhibits slightly better performance than \gls{CCDM} with \(n=3600\), while the performance gap between \gls{ESS} and \gls{CCDM}-200 remains approximately the same at around 0.19~bits/4D-sym. The performance of CCDM-3600 with respect to CCDM-200  (illustrated by the black arrows in \cref{fig:LPswp_GMI_Rloss}) reduces to 0.14~bit/4D. This can be explained by the fact that short blocklengths \gls{ESS}-200 and \gls{CCDM}-200 are more tolerant to nonlinear transmission impairments in comparison to the longer blocklength \gls{CCDM}-3600, and is also shown by the effective SNR \cite[Eq.~4]{Renner2017} in \cref{fig:LPswp_SNReff}. This observation has previously been shown in simulations \cite{Amari2019}. \gls{ESS} improves the performance in comparison with \gls{CCDM} at the same blocklength due to its lower rate loss, and this gain is independent from the launch power.

\Gls{BER} after shaping decoding is shown in \cref{fig:Dswp_BER_post_DM}. An outer \gls{HDFEC} staircase code with rate \SI{0.9373}~\cite{Smith2012} that corrects bit errors after LDPC decoding is assumed (see \cref{fig:shaping_block_diagram}). The \gls{BER} threshold is \(4.5\times10^{-3}\) \cite[Fig. 8]{Smith2012}, which makes \gls{ESS}-200 15\% better in reach (\SI{1610}{\kilo\meter}) compared to \gls{CCDM} (\SI{1400}{\kilo\meter}) with the same blocklength, which is an increase of \SI{210}{\kilo\meter}. Interestingly, \Gls{ESS}-200 is shown to slightly outperform \gls{CCDM}-3600. The combination of the rate of this staircase code together with a baudrate of 41.79~GBd and a net information rate of 9~bits/4D-sym results in a total data rate of just over 350~Gbit/s per channel.

\cref{fig:Dswp_GMI_Rloss} shows the AIR as a function of reach for the optimal launch power of 9.5~dBm. The AIR in \cref{fig:Dswp_GMI_Rloss} should be interpreted as the reach that an ideal PAS would achieve. For the considered rate (9~bit/4D-sym), ESS-200 offers approximately the same reach as CCDM-3600. \cref{fig:Dswp_GMI_Rloss} also shows that CCDM-200 reaches \SI{1880}{\kilo\meter}, while ESS-200 reaches \SI{2085}{\kilo\meter}. This corresponds to a gain of \SI{205}{\kilo\meter}. While the reach of the systems with non-ideal PAS are smaller (see results in \cref{fig:Dswp_BER_post_DM}), the reach increase offered by ESS-200 in comparison with CCDM-200 is approximately the same (\SI{210}{\kilo\meter} vs. \SI{205}{\kilo\meter}).

\vspace{-2mm}
\section{Conclusions}
Probabilistic \gls{ESS} is demonstrated for the first time in an optical transmission experiment. After \SI{1610}{\kilo\meter} of transmission over \gls{SSMF}, \gls{ESS} shows a 15\% reach increase over \gls{CCDM} at 350 Gbit/s per channel using dual-polarization \qam{64} symbols at a short blocklength regime. The performance of \gls{ESS} with a significantly lower blocklength is similar or better than that of long blocklength \gls{CCDM}. We believe that short blocklength ESS is a promising low-complexity shaping alternative to CCDM which could find application in next generation optical transceivers.

\vspace{-2mm}
\noindent\footnotesize\emph{\\
We acknowledge partial funding from the Dutch Netherlands Scientific Research Organisation (NWO) Gravitation Program on Research Center for Integrated Nanophotonics (Grant Number 024.002.033). The work of A. Alvarado is supported by the NWO via the VIDI Grant ICONIC (project number 15685) and has received funding from the European Research Council (ERC) under the European Union’s Horizon 2020 research and innovation programme (grant agreement No 757791). Fraunhofer HHI and ID Photonics are acknowledged for providing the Optical-Multi-Format Transmitter.}

\vspace{0mm}
\bibliographystyle{style/osajnl}
\bibliography{refshort}

\end{document}